\documentclass[12pt, preprint]{aastex}
\shorttitle{Spitzer IRAC Imaging of the AGB Stars in IC~1613}
\shortauthors{Jackson et al.}

\begin{document}

\def\HI{\ion{H}{1}}
\def\HII{\ion{H}{2}}
\def\msun{M_$\sun$}
\def\spitzer{$\it Spitzer$}
\def\micron{$\mu$m}
\def\solar{M$_\sun$}
\def\solaryr{M$_\sun$ yr$^{-1}$}
\def\ml{$\dot{M}$}

\title{A Spitzer IRAC Census of the Asymptotic Giant Branch
Populations in Local Group Dwarfs. II. IC~1613} 
\author{Dale C. Jackson, Evan D. Skillman, Robert D. Gehrz, 
Elisha Polomski, and Charles E. Woodward} 
\affil{Astronomy Department, University of Minnesota, 116
Church St. S.E., Minneapolis, MN 55419} 
\email{djackson@astro.umn.edu, skillman@astro.umn.edu, 
gehrz@astro.umn.edu, elwood@astro.umn.edu, chelsea@astro.umn.edu}


\begin{abstract} 
We present \spitzer\ {\it Space Telescope} IRAC photometry of the
Local Group dwarf irregular galaxy IC~1613. We compare our 3.6, 4.5,
5.8, and 8.0 \micron\ photometry with broadband optical photometry and
find that the optical data do not detect 43\% and misidentify an
additional 11\% of the total AGB population, likely because of
extinction caused by circumstellar material. Further, we find that a
narrowband optical carbon star study of IC~1613 detects 50\% of the
total AGB population and only considers 18\% of this population in
calculating the carbon to M-type AGB ratio. We derive an integrated
mass-loss rate from the AGB stars of 0.2-1.0$\times$10$^{-3}$
\solaryr\ and find that the distribution of bolometric luminosities
and mass-loss rates are consistent with those for other nearby
metal-poor galaxies.  Both the optical completeness fractions and
mass-loss rates in IC~1613 are very similar to those in the Local
Group dwarf irregular, WLM, which is expected given their similar
characteristics and evolutionary histories.
\end{abstract}



\section{Introduction}
IC~1613 is an isolated member of the Local Group and serves as the
prototype for the DDO Ir V type dwarf irregular (dI) galaxy
\citep{vdb00}. Discovered by \citet{wol06}, IC~1613's relative
proximity \citep[730 kpc;][]{dol01}, low inclination angle
\citep[i~=~38$\degr$][]{lak89}, favorable galactic latitude
(b~=~$-$60$\degr$), and low foreground reddening
\citep[E(B$-$V)=~0.025 mag;][]{sch98} make it an ideal target for
stellar population and interstellar medium (ISM) studies. This is
evidenced by the impressive number of previous studies of this galaxy,
summarized by \citet{vdb00}.

\citet{lak89} studied the neutral hydrogen in IC~1613. They found the
galaxy to be gas rich with the \HI\ distribution being lumpy in the
inner galaxy and smoother at larger radii. \citet{col99} and
\citet{ski03} performed deep photometry of single HST/WFPC2 fields in
the inner and outer galaxy, respectively. Both of these studies used
high-precision photometry to derive star formation histories. They
found continuous star formation over the past few hundred Myr, as well
as populations of intermediate age and ancient stars.

IC~1613 is, in most respects, a typical irregular galaxy. Its metal
abundance, current star formation rate, and dust content are all
normal for a galaxy of its luminosity. The morphology consists of a
prominent star forming region northeast of the galaxy center and a bar
running linearly through the galaxy center from southeast to
northwest. Perhaps the only exceptional property of IC~1613 is its
nearly complete absence of star clusters \citep{vdb79,wyd00}, which
may be evidence of its isolated evolution \citep{vdb00b}.

In this paper we present Part II of our \spitzer\ IRAC thermal
infrared study of the stellar populations in Local Group dwarf
irregulars with a census of the asymptotic giant branch stars in
IC~1613. In \S \ref{observations} we describe the observations, data
reduction, and photometry. In \S \ref{morphology} we present the
\spitzer\ IRAC images and describe the near- and mid-IR morphology of
IC~1613. We characterize the infrared color-magnitude diagram (CMD) in
\S \ref{photometry}. In \S \ref{AGB} we detail a complete census of
the asymptotic giant branch (AGB) stars, a comparison with broadband
optical photometry (\S \ref{complete}), a comparison with an optical
carbon star study (\S \ref{carbon}), the determination of mass-loss
rates (MLRs; \S \ref{mass_loss}), and the total luminosity
contribution from the AGB population (\S \ref{luminosity}). In \S
\ref{conclusions} we summarize our conclusions.


\section{Observations and Data Reduction}\label{observations}
\subsection{Infrared and Optical Data and Photometry}
\begin{figure}
\plotone{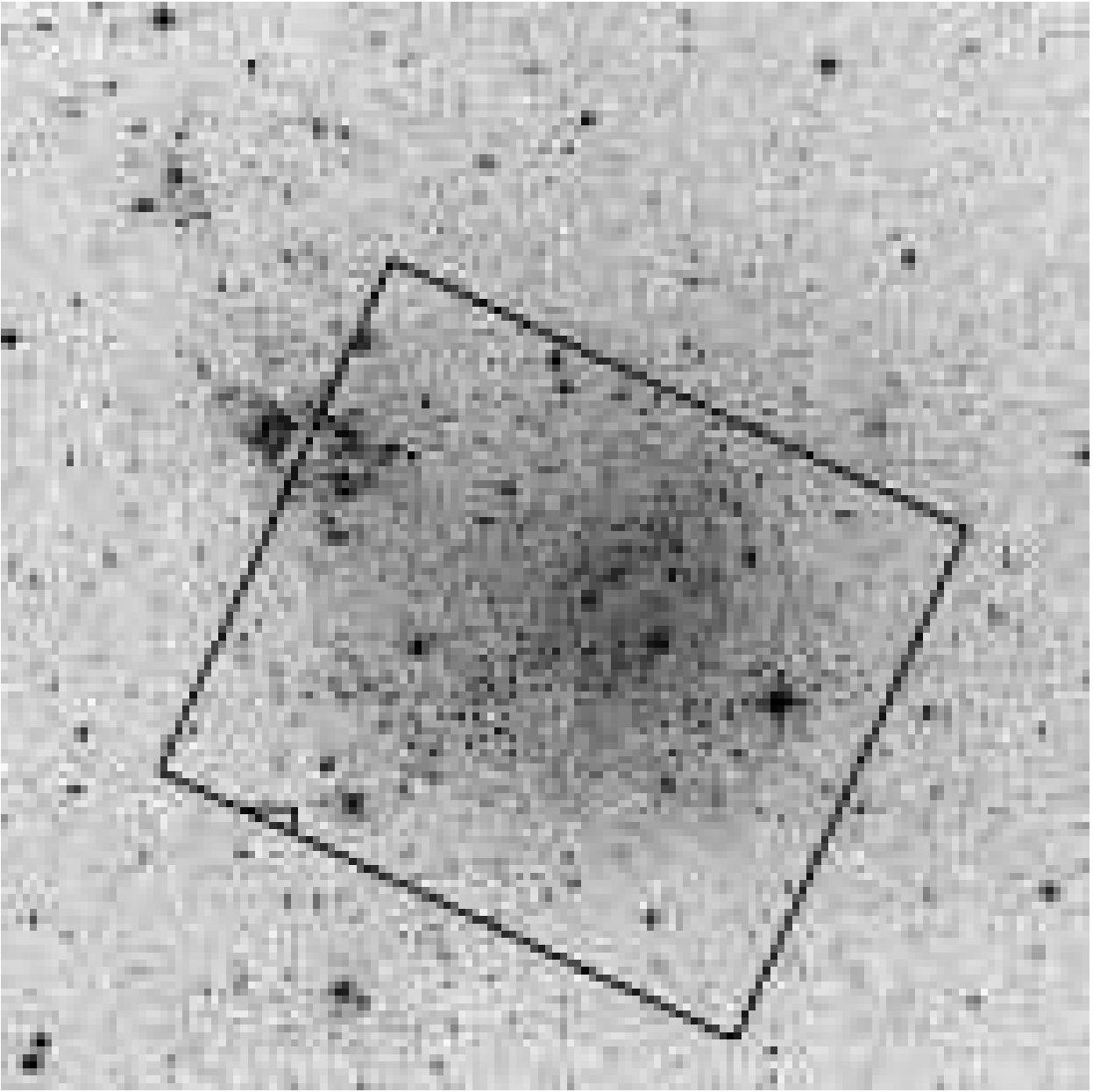}  
\caption{\label{coverage} Digitized Sky Survey image of IC~1613,
14$\arcmin$$\times$14$\arcmin$ in size. The region of sky with
coverage in all four IRAC bands is overlaid in black, and the small
triangular region with no optical {\it V} and {\it I} coverage is
shown in the southeast. North is up and East is left.}
\end{figure}

Images of IC~1613 were obtained with the Infrared Array Camera
\citep[IRAC;][]{faz04} on board the \spitzer\ {\it Space Telescope}
\citep{wer04} in two separate AORs on 2003 December 21 UT and 2005
December 23 UT (AORKEYs: r5051648 and r15892736), respectively. Both
AORs are part of a larger guaranteed time observing program (PID: 128,
PI: R.D. Gehrz). Five individual 200 s frames at 4.5 and 8.0 \micron\
and 32 individual 30 s frames at 3.6 and 5.8 \micron\ were used
resulting in total integration times of 1000 s per pointing at 4.5 and
8.0 \micron\ and 960 s per pointing at 3.6 and 5.8 \micron . These
data yielded 5-$\sigma$ point source sensitivities of 4.5, 3.5, 10.7,
and 16.5 $\mu$Jy at 3.6, 4.5, 5.8, and 8.0 \micron , respectively.
Because the IRAC 3.6 and 5.8 \micron\ field of view is offset from
that at 4.5 and 8.0 \micron, a 3$\times$2 mosaic in array coordinates
approximately 15$\arcmin$$\times$10$\arcmin$ for each channel was
created to map IC~1613. This resulted in a coverage area with data in
all four IRAC channels of approximately 9$\arcmin \times$9$\arcmin$ in
size, centered at $\alpha$(2000)=02:06:30,
$\delta$(2000)=01:04:50. Figure \ref{coverage} shows this field of
view overlaid on the Digitized Sky Survey image of IC~1613.

IRAC data (pipeline version 13.2.0) were reduced with the
MOPEX\footnote{MOPEX is available from the Spitzer Science Center at
http://ssc.spitzer.caltech.edu/postbcd/} reduction package, version
2006 March 1. The backgrounds of individual frames were matched using
the overlap routine. The mosaic program was used for outlier
detection, image interpolation and co-addition, and to create the
final mosaics.  Very bright objects imaged in the program immediately
preceding our observations resulted in persistent images in the 8.0
\micron\ data taken in 2003. These artifacts were effectively removed
by subtracting a median-combined image of all of the dithered frames
from each individual frame.

Point source photometry was done using the DAOPHOT II photometry
package \citep{ste87}. A sharpness clip was applied to the photometry
in each of the IRAC bands. This clip removes any objects from the
photometry list whose radial profiles are significantly steeper or
broader than the profile of the derived point spread function
(PSF). We adopt the standard DAOPHOT errors as the total photometric
errors for our sources. The IRAC instrument also has an estimated
absolute calibration accuracy of 3\% \citep{rea05}, although we do not
include this uncertainty in our photometric errors.

Optical {\it V} and {\it I} photometry of IC~1613 were obtained from
the OGLE-II microlensing survey \citep{uda01}. These data are used to
aid in identifying the stellar type of the stars detected in our IRAC
data, which is challenging with IRAC data alone (see \S
\ref{photometry}). The optical data were acquired in the context of
detecting objects that are variable (due to either microlensing events
or intrinsic stellar variability) by taking a large number of frames,
each with relatively short integration times. The photometry presented
here is the mean of the individual photometric observations, including
those objects found to be variable. The field of view of the
\citet{uda01} optical data covers the entire IRAC 3.6, 4.5, 5.8, and
8.0 \micron\ coverage area, except for a small triangular region
$\sim$0.9 square arcmin in size in the south-eastern edge of the IRAC
coverage area (see Figure \ref{coverage}).

\subsection{Foreground Star Contamination}
To estimate the number of foreground stars in both the optical and IR
data we used the Milky Way stellar population synthesis model of
\citet{rob03}. Sources were modeled for the Galactic coordinates of
IC~1613 in a one square degree field to reduce statistical error. The
\citet{rob03} model provides magnitudes in the Johnson-Cousins L-band,
whose central wavelength is close to that of the IRAC 3.6 \micron\
band. In the region of IC~1613 where we have coverage in the optical
and all four IRAC channels, we expect 15 foreground stars in our IR
color-magnitude diagram (CMD). These 15 foreground stars are expected
to have $-$7.5$<$M$_{3.6 \mu m}<-$16 and [3.6]$-$[4.5] colors near
zero.


\section{Infrared Galaxy Morphology}\label{morphology}
\begin{figure}
\plotone{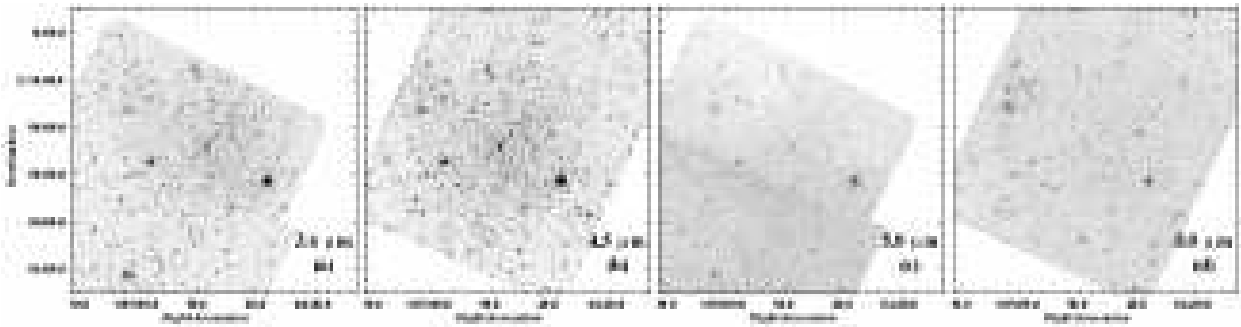}  
\caption{\label{irac} IRAC 3.6, 4.5, 5.8, and 8.0 \micron\ images of
IC~1613. North is up and East is left.}
\end{figure}

IRAC 3.6, 4.5, 5.8, and 8.0 \micron\ images of IC~1613 are shown in
Figure \ref{irac}. In general, the stars in these images are uniformly
distributed throughout the images, consistent with the regular
elliptical distribution of red giants discussed by \citet{baa63}, and
dubbed Baade's Population II `sheet' by \citet{san71}. In addition to
the smooth distribution of stars, a concentration of bright stars is
seen in the large star forming complex in the northeast. The only
diffuse emission detected in these images is 8.0 \micron\ emission
from hot ($\sim$350 K) dust and/or polycyclic aromatic
hydrocarbons. This emission is detected in regions roughly coincident
with strong H$\alpha$ emission \citep[][and references
therein]{jac06}.


\section{Comparison of Optical and IR Photometry}\label{photometry}
\begin{figure}
\plotone{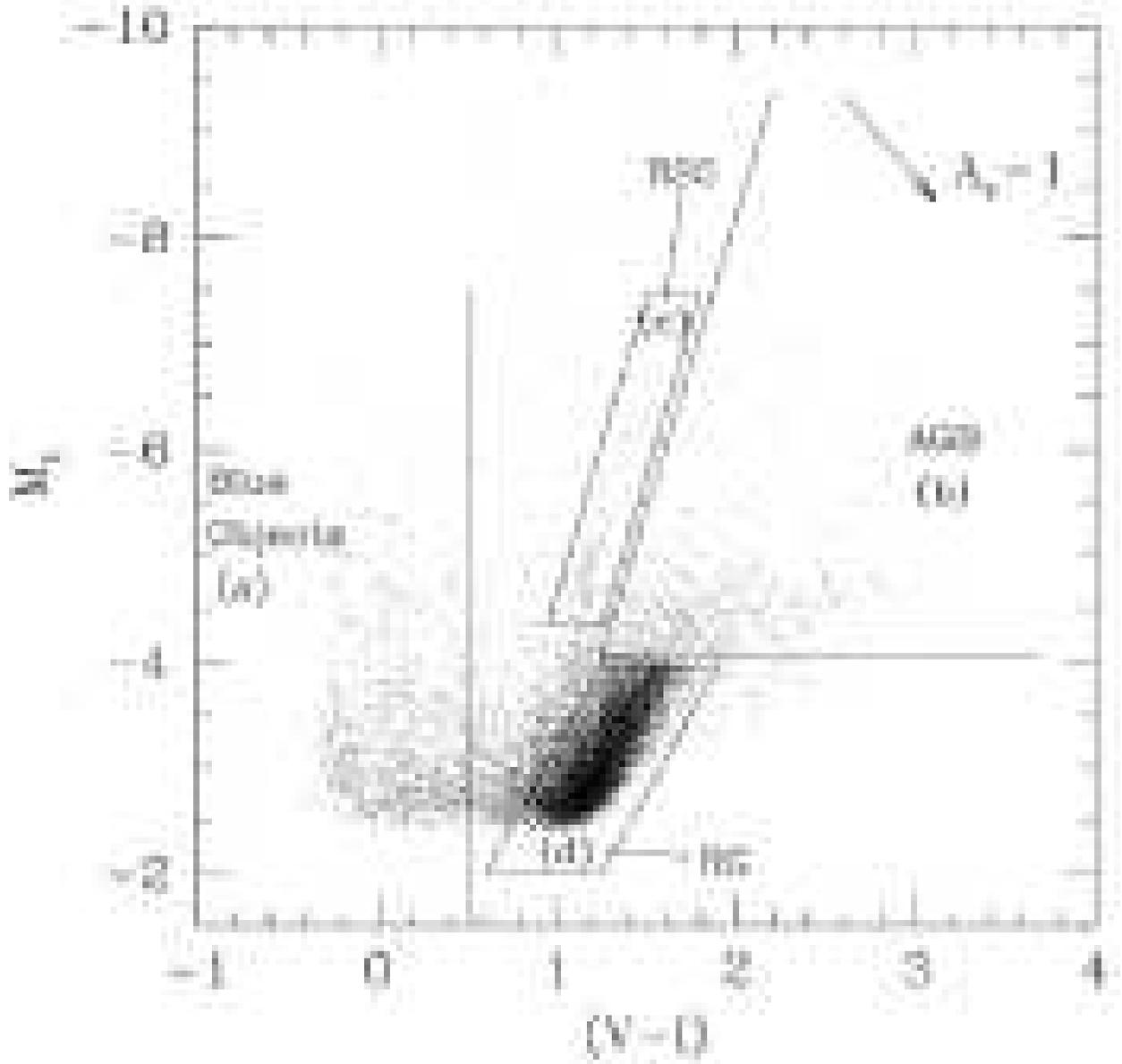}  
\caption{\label{optical_cmd} Absolute I versus {\it V}$-${\it I} color
magnitude diagram of IC~1613 using data from \citet{uda01}. The
diagram is segregated into regions containing (a) blue objects, (b)
AGB stars, (c) red supergiants, and (d) sub-TRGB red giants to aid in
determining the stellar types of stars in the infrared color-magnitude
diagram (Figure \ref{ir_cmd}). The TRGB is located at M$_I$~=~$-$4, in
the center of the gap separating regions b and d.}
\end{figure}

The optical color-magnitude diagram (CMD) from the \citet{uda01} data
is shown in Figure \ref{optical_cmd}, segregated into regions
consisting of (a) blue objects that can include massive main sequence
stars and unresolved \HII\ regions, (b) asymptotic giant branch (AGB)
stars, (c) red supergiants (RSGs), and (d) red giants (RGs) below the
tip of the red giant branch (TRGB). We refer hereafter to the stars
called out in (d) as sub-TRGB stars. These regions are used to
identify the stellar populations in the infrared CMD, since IRAC data
alone provide little information on stellar effective temperatures.
These indicated regions are separated by gaps reflecting the
estimated 1-$\sigma$ photometric error, which were employed to minimize
misidentification of stellar types solely due to photometric errors.
Reddening from circumstellar and interstellar material can also cause
such misidentification. The optical CMD shows features of both recent
and very old star formation, consistent with the finding of
\citet{ski03} that IC~1613 has a relatively constant star formation
rate over the age of the Universe.

\begin{figure}
\plotone{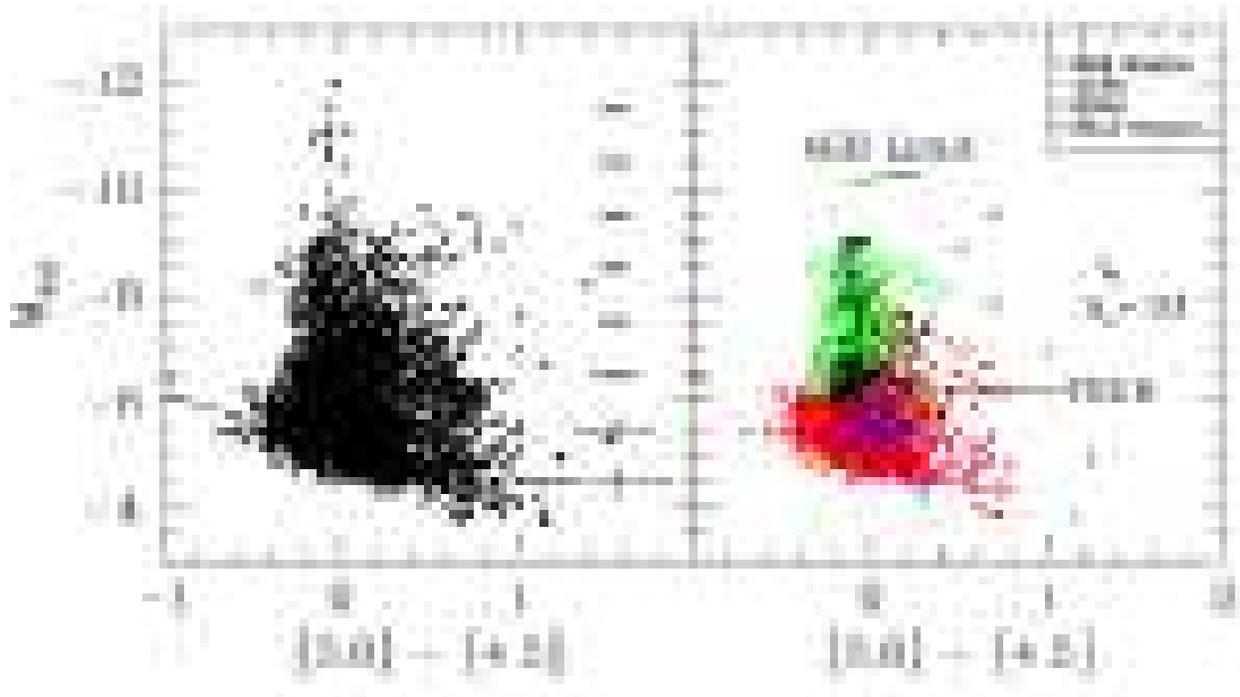}  
\caption{\label{ir_cmd} 3.6 \micron\ absolute magnitude versus
[3.6]$-$[4.5] color magnitude diagram of IC~1613. The left panel shows
all of the point sources detected at 3.6 and 4.5 \micron . The right
panel shows objects also detected in the \citet{uda01} {\it V} and
{\it I} data, segregated according to Figure \ref{optical_cmd} with
sub-TRGB red giants as red squares, AGB stars as green triangles, RSGs
as black stars, and blue objects as blue circles. Photometric
1-$\sigma$ errors, averaged over 1 mag bins, are shown on the right
side of the left panel. The AGB limit, TRGB, and a vector showing 10
magnitudes of visual extinction are shown in the right panel.  See \S
\ref{photometry} for discussion.}
\end{figure}

Figure \ref{ir_cmd} shows the 3.6 \micron\ absolute magnitude versus
[3.6]$-$[4.5] \micron\ CMD. The left panel shows all of the stars
detected at both 3.6 and 4.5 \micron , with photometric errors,
averaged over one magnitude bins, displayed. The left panel also shows
the 50\% completeness limit (black line) derived from artificial star
tests. The right panel shows the stars detected in {\it V}, {\it I},
and at 3.6 and 4.5 \micron , with RSGs shown as black stars, AGB stars
as green triangles, red giants as red squares, and blue objects as
blue circles. The right panel contains a vector showing the
displacement a star would experience due to 10 visual magnitudes of
extinction and reddening \citep{ind05, rie85}, though this vector does
not include the effects of dust emission, which can be significant in
the mid-IR. The right panel also shows the AGB limit, which was
determined by taking the \citet{gro06} models for carbon-rich AGB
stars with T$_{eff}$~=~2650 and T$_{eff}$~=~3600 and negligible
mass-loss and scaling them to an absolute bolometric magnitude of
M$_{bol}$~=~$-$7.1. The AGB limit depicted in this figure is a line
connecting these two models. AGB stars that are losing significant
mass can be brighter than this limit due to thermal emission of the
mass-loss material increasing their 3.6 \micron\ flux (as shown in
detail in \S \ref{mass_loss}), however, any such objects would also
have red [3.6]$-$[4.5] colors. None of these objects are observed,
thus, we conclude all of the stars brighter than the AGB limit are true
red supergiants.

\begin{figure}
\plotone{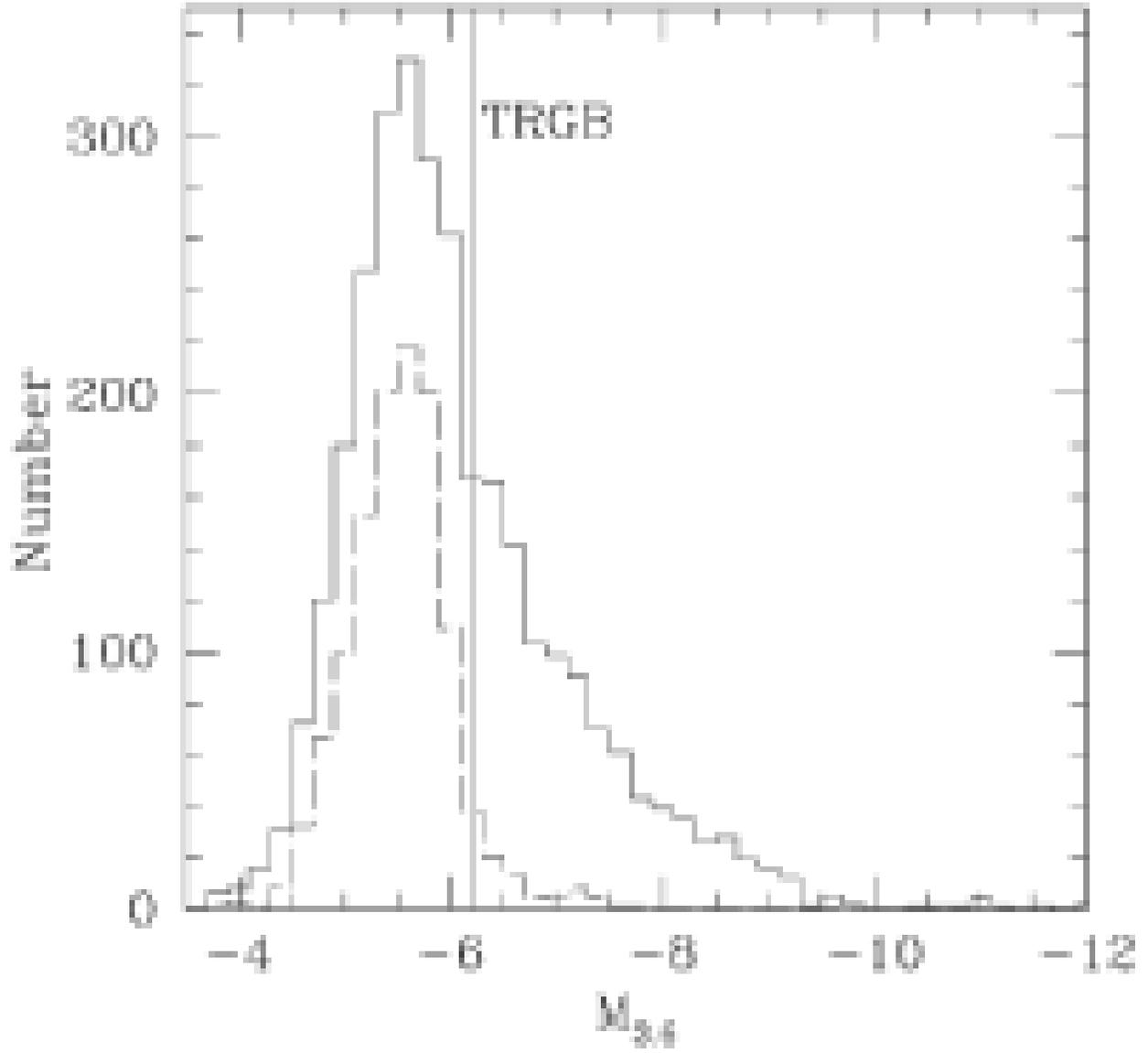}  
\caption{\label{lum_fn} The 3.6 \micron\ luminosity functions for all
of the stars in IC~1613 that were detected at 3.6 and 4.5 \micron
({\it solid line}) and for those detected at 3.6 and 4.5 \micron\ and
optically classified as sub-TRGB red giants ({\it dashed line}). The
value we adopt for the 3.6 \micron\ TRGB is shown at M$_{3.6}$=$-$6.2
(see \S \ref{photometry} for discussion).}
\end{figure}

The right panel also shows the 3.6 \micron\ TRGB. The value of the
TRGB we adopt (M$_{3.6}$=$-$6.2$\pm$0.2) was determined by inspecting
the 3.6 \micron\ luminosity function for sources detected at both 3.6
and 4.5 \micron . The 3.6 \micron\ luminosity function (the number of
stars in each 0.2 magnitude bin at 3.6 \micron ) is shown in Figure
\ref{lum_fn}. We adopt this value of the TRGB based on the abrupt drop
in detections at that magnitude. Also, optically classified sub-TRGB
red giants with blue [3.6]$-$[4.5] colors are observed with M$_{3.6}$
magnitudes up to this value, but above this value only optically
classified sub-TRGB red giants with very red colors are observed.  The
IR fluxes of these objects are consistent with mass-losing AGBs (see
\S \ref{AGB}) rather than sub-TRGB red giants, supporting our adopted
value of M$_{3.6}$=$-$6.2 for the 3.6 \micron\ TRGB.  This value is
0.2 magnitudes fainter than the value found for the Large Magellanic
Cloud (LMC) \citep[M$_{L^\prime}$=$-$6.4;][]{van05} and 0.4 magnitudes
lower than that adopted for WLM \citep{jac07}. It is unlikely that the
difference in values of the TRGB between IC~1613 and WLM are actually
so disparate, given their similar metallicities. The difference in
these values likely reflects the uncertainty in our adopted TRGB
values. This uncertainty is not a major concern, however, because
shifting the value of the TRGB by up to 0.5 magnitudes affects the
detection statistics only slightly (see \S \ref{complete}).

In general, the M$_{3.6}$ versus [3.6]$-$[4.5] CMD of IC~1613 is very
similar to that of WLM, with the only major difference being that
there are significantly more sub-TRGB red giants detected in IC~1613
than in WLM because WLM is more distant. As is the case of WLM, a
small population of stars redward of the main stellar distribution is
detected, though fewer are seen in IC~1613 than in WLM. As we discuss
in \S \ref{mass_loss}, the infrared fluxes and colors of these objects
are consistent with mass-losing AGB stars.

\begin{figure}
\plotone{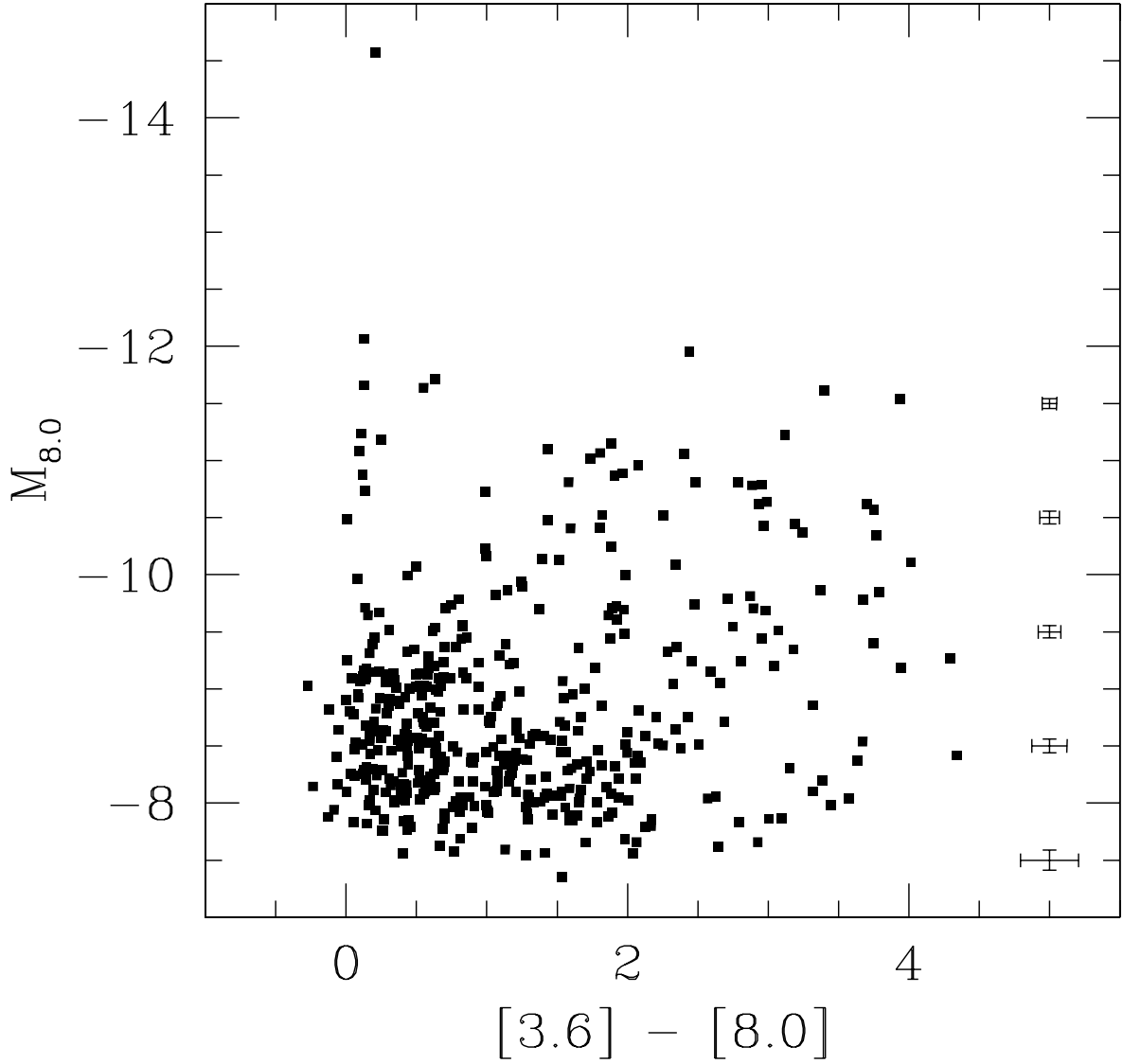}  
\caption{\label{ch4cmd} The 8.0 \micron\ absolute magnitude versus
[3.6]$-$[8.0] color magnitude diagram of IC~1613. Photometric
1-$\sigma$ errors, averaged over 1 mag bins are shown on the right.}
\end{figure}

Figure \ref{ch4cmd} is the M$_{8.0}$ versus [3.6]$-$[8.0] CMD for
IC~1613. This CMD shows a narrow vertical feature with
[3.6]$-$[8.0]$\sim$0 and another broad distribution of red objects
with $-$12~$<$~M$_{8.0}$~$<$~$-$7.5 1~$<$~[3.6]$-$[8.0]~$<$~4. These
red objects are also the most luminous, reddest objects in the
M$_{3.6}$ versus [3.6]$-$[4.5] CMD, with typical colors between 0.5
and 1.0.

\begin{figure}
\plotone{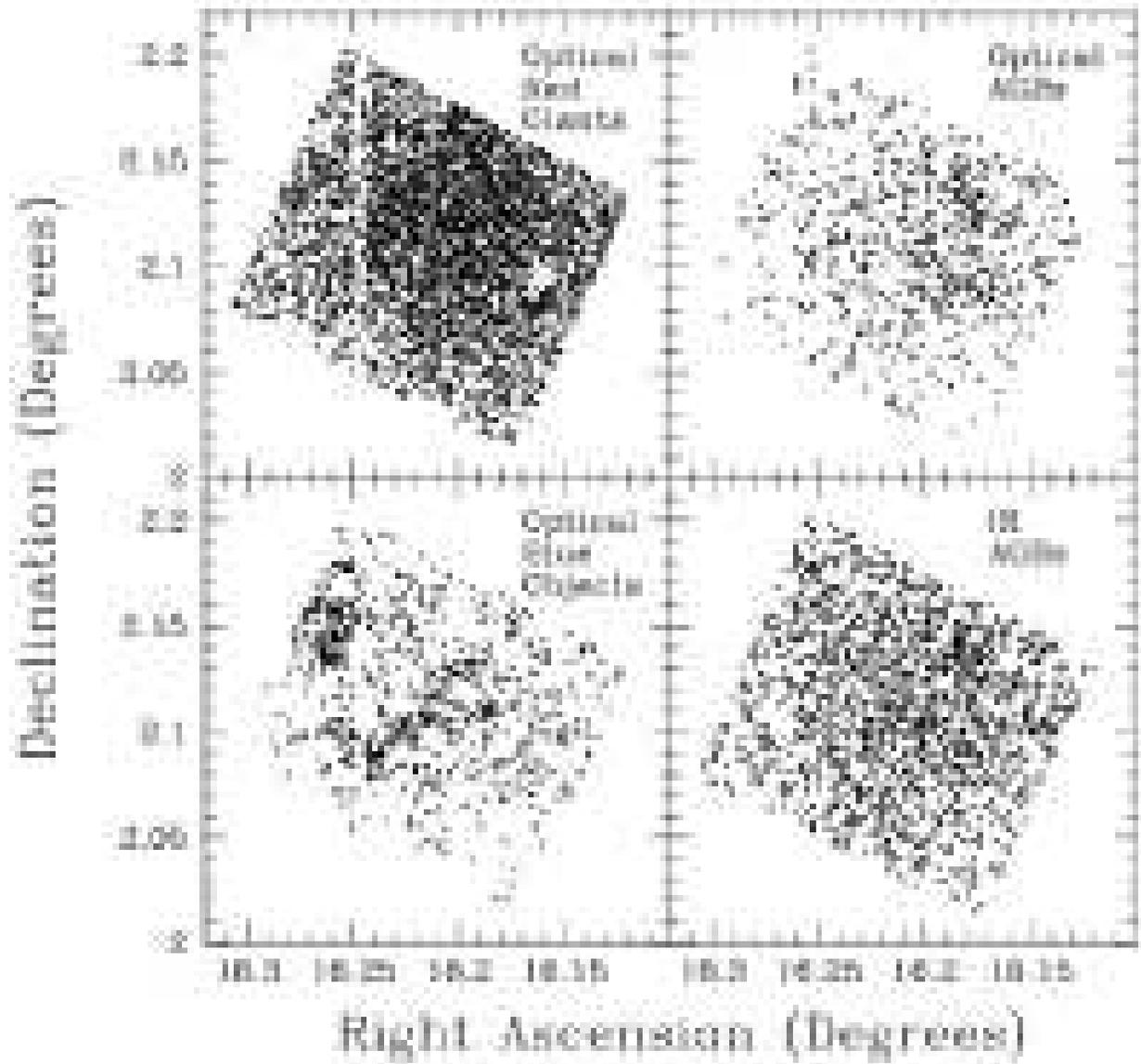}  
\caption{\label{xy} Positions of stars according to their optical
classification, with the exception of the bottom right panel, which
shows the positions of AGB stars detected in the IR. }
\end{figure}

Figure \ref{xy} shows the spatial distributions of different stellar
types based on their optical and IR fluxes. The distributions of both
red giants and AGB stars are very smooth with no obvious
concentrations aside from the radial stellar gradient. A thin gap is
observed in the optical data, which is certainly an instrumental
effect, though this is not mentioned in \citet{uda01}. There are two
conspicuous features in the blue objects, which outline the young
stellar distribution; a bar running across the midplane of the galaxy
and a large clump in the northeast, which highlights the most active
star forming complex in IC~1613. It is clear from this figure that
neither the optical nor the IR data presented here reach the outermost
regions of the stellar population, as the stellar density is still
dropping off at the edges of the images. In the analogous plot in
\citet{jac07}, the effects of crowding could clearly be seen in WLM as
clumps of bright, young stars coincided spatially with an absence of
fainter red giants. In IC~1613 we see no evidence for such an effect,
likely because IC~1613 is seen much more face on \citep[{\it
i}~=~38$\degr$,][]{lak89} than WLM \citep[{\it
i}~=~69$\degr$,][]{jac04}.

In Table \ref{stats} we present the detection statistics for the
optical and infrared photometry of IC~1613. We detect 5211, 3162, 886,
and 618 point sources at 3.6, 4.5, 5.8, and 8.0 \micron ,
respectively, and the \citet{uda01} {\it V} and {\it I} data contain
7266 point sources within the optical and IR coverage region. We
detect 183 point sources in all four IRAC channels with no corresponding
detection in the optical. Three of these are RSGs (i.e., they are
above the AGB limit), two of these are very red objects below the
TRGB, and the remaining 178 are either RSGs, AGB stars, or blue
objects.


\section{The AGB Stars}\label{AGB}
As shown in Figure \ref{ir_cmd}, our completeness limit is more than
0.5 mag fainter than the TRGB for the bluest objects and 1.5 mag
fainter than the TRGB for red objects. Consequently, our data
constitutes a complete census of the super-TRGB AGB stars with the
exception of a very small number of objects not detected due to
crowding.

As described in \citet{jac07}, though our data should represent a
complete census of the AGB stars, we cannot distinguish between AGB
stars, RSGs, and blue objects using IRAC data alone.  However, above
the TRGB, optically identified RSGs and blue objects comprise only 9\%
(95 out of 1041) of the objects between the TRGB and the AGB
limit. Therefore, we assume any object between the TRGB and the AGB
limit is an AGB star unless its optical identification dictates
otherwise, with the understanding that there may be a small amount of
contamination by other stellar types.

\subsection{Optical Completeness}\label{complete}
Altogether we detect 1052 AGB stars in our IRAC data. Of these 57\%
are detected in the \citet{uda01} {\it V} and {\it I} data. In
addition to the 43\% of AGB stars detected in the IR but not seen in
the optical, 11\% of the IR detected AGB stars are located in the
optical CMD where they would be misidentified as sub-TRGB red giants.

For comparison, in WLM \citet{jac07} found 39\% of the IR detected AGB
stars were not seen in the optical data of \citet{mas06}, and an
additional 4\% were optically detected, but misidentified as sub-TRGB
red giants.  An important check of our AGB statistics for WLM was
recently provided by \citet{val07}, who performed J and K$_S$ near-IR
photometry. Altogether they detect 355 AGB stars (carbon~+~M-type
AGBs), in comparison with the 331 AGB stars we detect in the same
field of view. It is not surprising that the number of AGBs we detect
in WLM and IC~1613 are in such close agreement given the striking
similarities in the evolution, and particularly in the intermediate
age star formation histories, of these galaxies \citep[Figures 25 and
35 in][]{dol05}.
 
\begin{figure}
\plotone{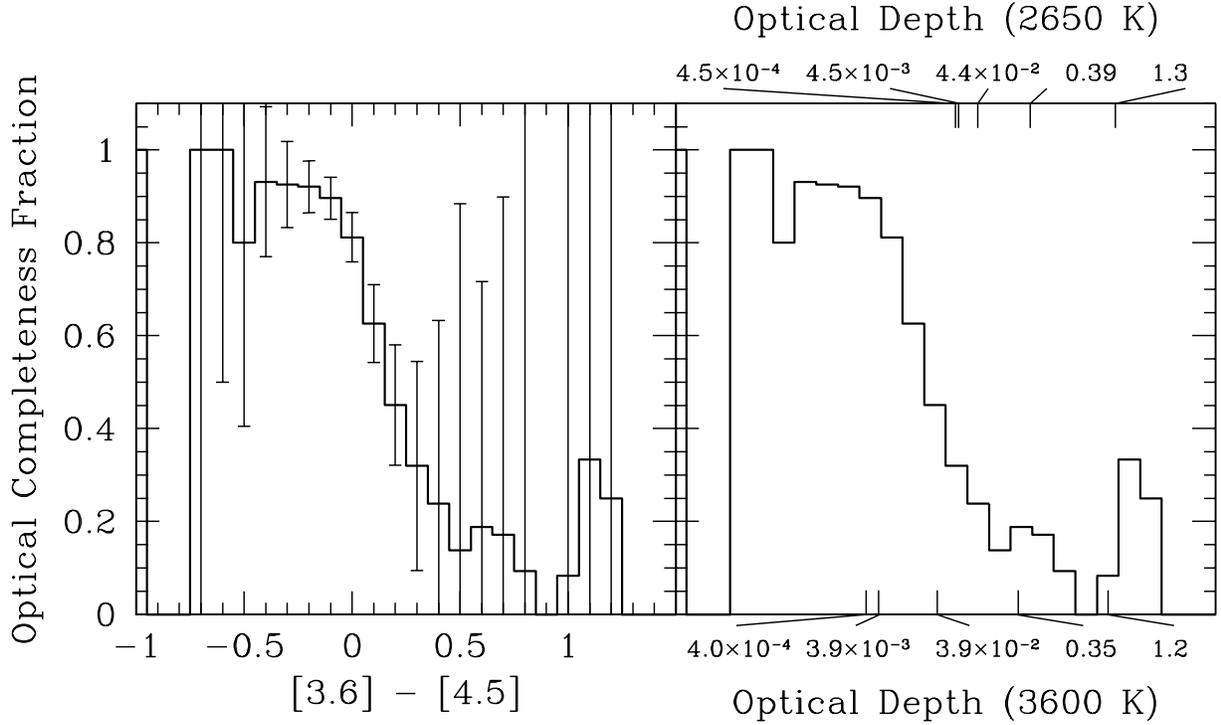}  
\caption{\label{frac} Fraction of objects more luminous than the TRGB
at 3.6 \micron\ that are also detected in the \citet{uda01} {\it V}
and {\it I} photometry as a function of [3.6]$-$[4.5] color ({\it left
panel}) and as a function of wind optical depth assuming AGB stars
with wind composition of 85\% AMC + 15\% SiC and effective
temperatures of 2650 K ({\it top axis}) and 3600 K ({\it bottom
axis}). See \S \ref{mass_loss} for a detailed discussion of mass-loss
modeling. The error bars were created by taking the square root of the
number of optically detected AGBs and dividing by the total number of
AGBs seen in at 3.6 and 4.5 \micron\ and therefore only represent the
degree to which small number statistics might affect the overall
trend.}
\end{figure}

In Figure \ref{frac} we show the fraction of optical completeness as a
function of [3.6]$-$[4.5] color (left panel) and as a function of the
modeled AGB wind optical depth (right panel; AGB mass-loss modeling is
described in detail in \S \ref{mass_loss}). This figure shows the
optical completeness fraction is a smoothly declining function of
[3.6]$-$[4.5] color, as is expected if the incompleteness is due to
extinction by circumstellar material. Much like in the case of WLM, we
see the completeness fraction fall to 0\% at a [3.6]$-$[4.5] color of
$\sim$0.9.

It is important to understand to what degree our adopted value of the
TRGB affects these completeness statistics. By shifting our adopted
value 0.5 mag fainter the fraction of optically undetected AGB stars
changes by only 2\%, while shifting the TRGB 0.5 mag brighter would
increase the optically undetected AGB fraction by 10\%. As in the case
of WLM, this increase in optical incompleteness with increasing
luminosity supports our conclusion that the optically undetected AGB
stars are not seen because they are enshrouded by circumstellar
material, since the mass-loss phase in AGB stars occurs somewhat above
the TRGB and should consequently be more conspicuous at higher
luminosities.

\subsection{The Carbon Stars}\label{carbon}
Narrowband and broadband optical photometry of IC~1613 using the
CN/TiO technique was performed by \citet{alb00}. In comparison with
our IRAC photometry, the \citet{alb00} data detected 50\% of the total
AGB population seen in our IRAC data. This value is slightly lower
than the fraction detected by \citet{uda01} and is caused by the
brighter completeness limits of the narrowband data, since the
broadband data from \citet{alb00} is significantly more sensitive than
that of \citet{uda01}.

\citet{alb00} distinguish between oxygen- and carbon-rich AGBs using a
color-color diagram. At relatively red {\it (R-I)} colors oxygen- and
carbon-rich AGBs have easily distinguishable CN-TiO colors, however,
AGBs with bluer {\it (R-I)} colors are not easily separated.
Consequently, a color limit is imposed, redward of which the
discrimination of chemical composition can be made. Within our IRAC
coverage area \citet{alb00} identified 74 carbon stars. Assuming their
C/M ratio of 0.64, this yields 190 total AGB stars (C~+~M-type). Thus,
after applying their imposed color limit, 18\% of the total AGB
population was considered for the C/M ratio.

\begin{figure}
\plotone{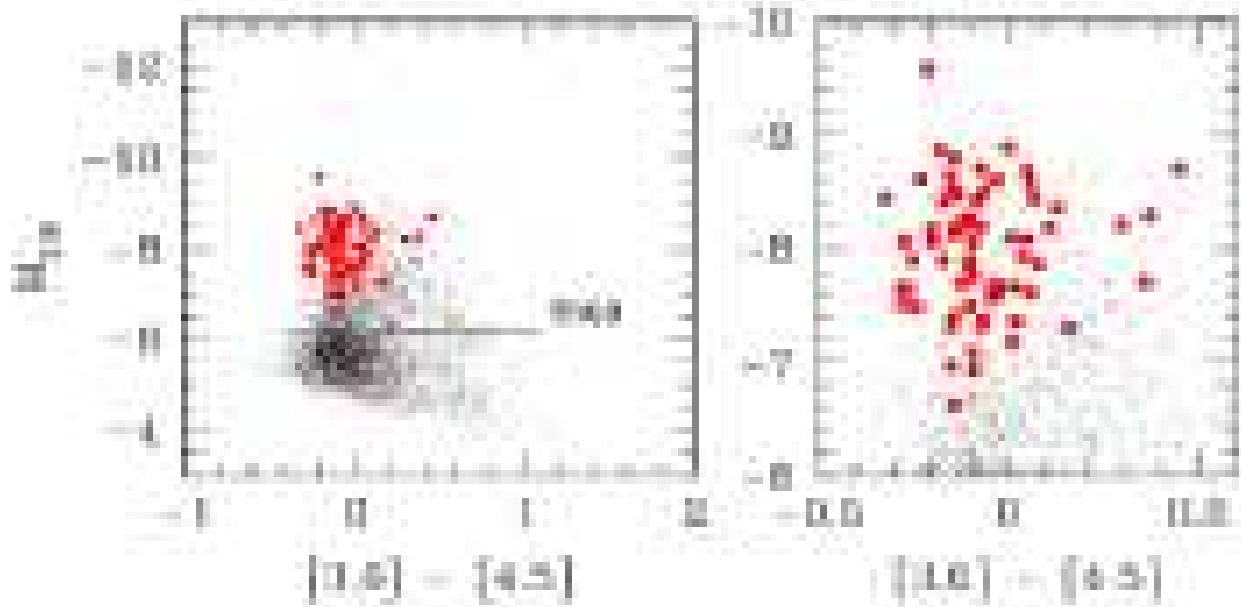}  
\caption{\label{carbon_cmd} Absolute 3.6 \micron\ versus [3.6]$-$[4.5]
color-magnitude diagram. The small black points are our IRAC
detections and the large red circles are our measurements of the
carbon stars identified by \citet{alb00}. Note in particular the four
carbon stars with [3.6]$-$[4.5]~$>$~0.2, which are likely losing
significant mass. The right panel is magnified to help distinguish the
points.}
\end{figure}

Figure \ref{carbon_cmd} shows our IRAC photometry of the carbon stars
identified by \citet{alb00}. The carbon stars have relatively blue
colors, probably due to the low-metallicity, and consequently the low
optical depth of circumstellar wind material of AGBs in IC~1613,
though four red carbon stars are seen that are likely losing
significant mass. There are also three carbon stars identified by
\citet{alb00} that we find to be well below the TRGB at 3.6
\micron. These three stars are also below the TRGB in the optical, and
are possibly extrinsic carbon stars. Because it is very unlikely that
these three objects are in fact true carbon-rich AGB stars we do not
consider them in any of the carbon star statistics. With the exception
of these three objects, the carbon stars identified by \citet{alb00}
have relatively bright 3.6 \micron\ luminosities compared to the rest
of the AGB stars.

The comparison between our IR photometry and the narrowband optical
data in IC~1613 is nearly identical to the values we found for this
same comparison in WLM \citep{jac07}. In WLM we found the completeness
fraction of the \citet{bat04} optical narrowband data was 63\% of the
total AGB population observed in our IRAC data and the C/M ratio was
based on 18\% of the IRAC detected AGB stars. Like in the comparison
of broadband optical completeness, the striking similarity between AGB
populations in IC~1613 and WLM is likely due to their similar star
formation histories and metallicity evolution.

In \citet{jac07} we stated that the \citet{bat04} optical CN/TiO study
of WLM {\it detected} only 18\% of the total AGB population, which is
misleading. Again, because of the necessity to impose a color limit to
discriminate between carbon- and oxygen-rich AGBs, relatively blue AGB
stars are ignored in the calculation of the C/M ratio. In WLM 18\% of
the total AGB population was used to calculate the C/M ratio, while an
additional 45\% of the total population was detected but not used in
the C/M ratio because the chemical composition could not be
determined. This point is critical to consider when comparing C/M
ratios obtained using different methods. As an example, recent near-IR
photometry of WLM \citep{val07} found a C/M ratio of 0.58 (corrected
for foreground contamination). This is in sharp contrast with the
\citet{bat04} value of 12.4, though these values must not be directly
compared because, as is shown here, the \citet{bat04} C/M ratio was
calculated using 18\% of the AGB stars, while the \citep{val07} value
considered the entire AGB population.

\subsection{Mass-Loss Rates}\label{mass_loss}
The mass-loss rates (MLRs) of AGB stars in IC~1613 were determined by
comparing their infrared fluxes with the radiative transfer models of
\citet{gro06}. The [3.6]$-$[4.5] color of AGB stars corresponds to a
unique wind optical depth ($\tau$), given assumptions about the wind
composition and stellar effective temperature (T$_{eff}$), as we
discuss below. The [3.6]$-$[4.5] colors were linearly interpolated
onto the \citet{gro06} models to estimate $\tau$, then scaled
according to the \citet{van00} MLR prescription, where:
$\dot{M}$~$\propto$~$\tau$$\psi^{-\frac{1}{2}}$L$^{\frac{3}{4}}$,
$\psi$ is the dust-to-gas ratio, and L is the stellar luminosity. For
IC~1613 we adopt the nebular oxygen abundance of 12+log(O/H)~=~7.62
\citep{lee03}, which corresponds to a metallicity of 0.9\%
Z$_{\sun}$. Given that the dust-to-gas ratio scales as
$\psi$~=~$\psi$$_{\sun}$10$^{-[Fe/H]}$ and $\psi$$_{\sun}$~=~0.005
\citep{van05}, we adopt $\psi$$_{IC~1613}$~=~4.8$\times$10$^{-4}$.

It is more appropriate to apply the metallicity of IC~1613 as it was a
few Gyr ago when the current AGB population was forming, rather than
the current metallicity. However, determining this value empirically
is difficult, and \citet{ski03} find very little metallicity evolution
in IC~1613 over the past few Gyr, so applying the current metallicity
to the AGB stars should be reasonably accurate.

There are two major uncertainties in deriving MLRs for AGB stars using
only our 3.6 and 4.5 micron data. The first is that there is a
degeneracy between $\tau$ and T$_{eff}$, such that cool AGB stars
losing no mass can have the same [3.6]$-$[4.5] color as warmer AGBs
that are losing mass. Because of this degeneracy we are forced to
derive MLRs by adopting a single T$_{eff}$ for all of the AGB stars,
which is certainly not physical. As a consequence, the MLRs we find
for individual blue AGB stars (i.e., AGB stars with
[3.6]$-$[4.5]~$<$~0.5) should not be over-interpreted. \citet{mar06}
showed that atmospheric molecular absorption features can also effect
the IRAC colors of AGB stars, introducing additional
uncertainty to these measurements, although these effects have not yet
been quantified. However, as emphasized in \citet{jac07}, because the
majority of the integrated mass-loss for IC~1613 comes from sources
much redder than [3.6]$-$[4.5]~$>$~0.5, where the MLR we calculate is
much less sensitive to T$_{eff}$, the integrated MLR for the galaxy as
well as those found for red AGB stars are robust.

The second major uncertainty in our MLR calculation is that we have no
{\it a priori} information about the chemical composition of the AGB
winds, and therefore, must assume a single chemical composition for
the entire AGB population. This assumption is also unphysical, and the
disparity between the integrated MLRs we measure for all of the
modeled wind compositions and effective temperatures differ by a
factor of four.

\begin{figure}
\plotone{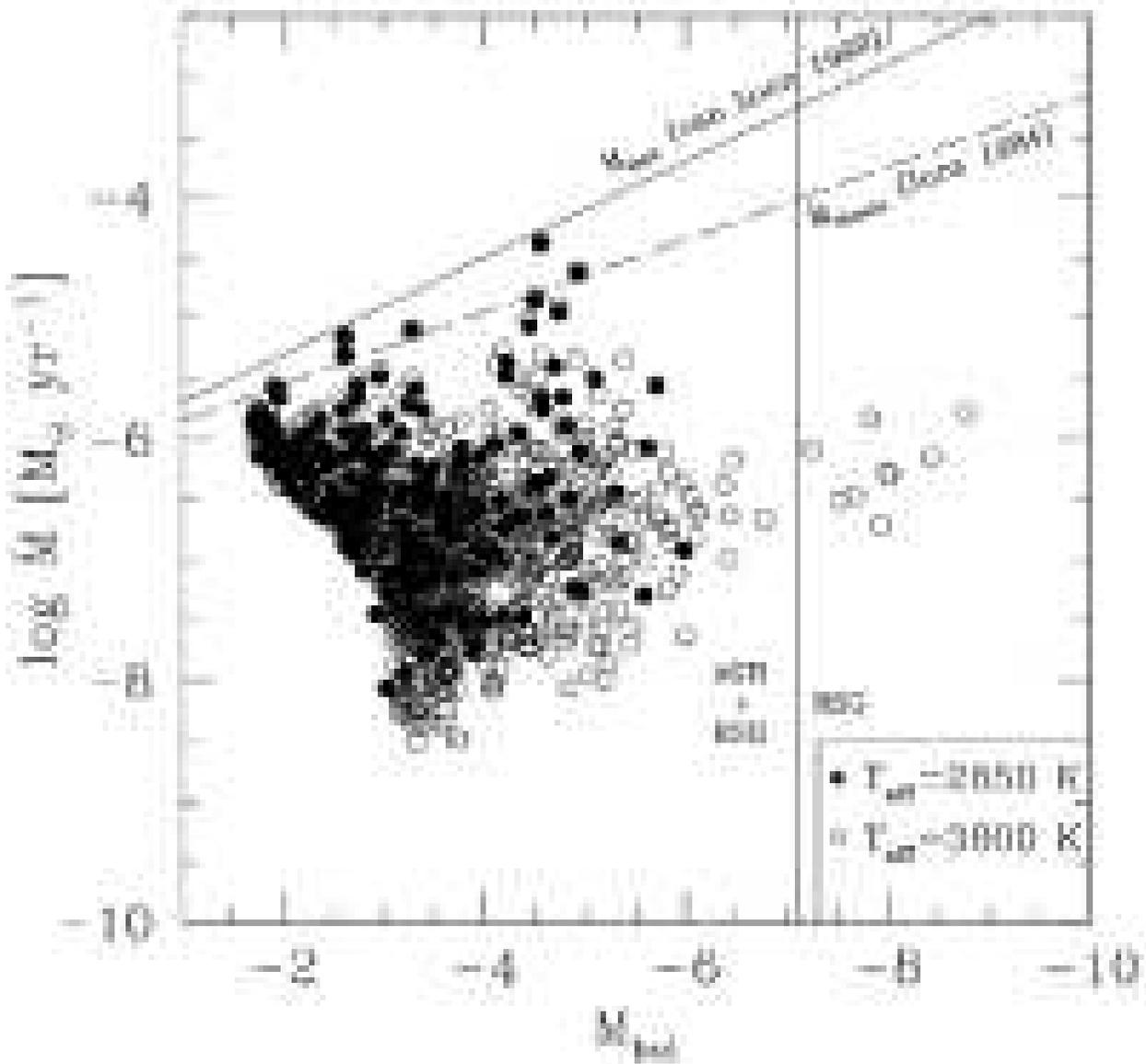}
\caption{\label{mbol} The mass-loss rate versus bolometric magnitude
for all objects brighter than the the TRGB at 3.6 \micron . The
mass-loss rates assume a gas-to-dust ratio of 4.8$\times$10$^{-4}$, a
wind composition of 85\% AMC + 15\% SiC and effective temperatures of
T$_{eff}$~=~2650 K ({\it filled} circles) and T$_{eff}$~=~3600 K ({\it
open} circles), using the mass loss prescription of \citet{van00} (see
\S \ref{mass_loss}).  The bottom dashed line is the \citet{jur84}
single-scattering mass-loss limit, and the top solid line is the
empirical maximum mass-loss limit observed in the LMC
\citep{van99}. The AGB limit is shown as a vertical line at
M$_{bol}$~=~$-$7.1.}
\end{figure}

Figure \ref{mbol} shows the MLR versus bolometric luminosity for each
AGB and RSG in IC~1613 assuming a wind composition of 85\% amorphos
carbon (AMC) + 15\% SiC and effective temperatures of T$_{eff}$~=~2650
K ({\it filled} circles) and T$_{eff}$~=~3600 K ({\it open}
circles). Again, we consider any object detected at 3.6 and 4.5
\micron\ that is above the TRGB at 3.6 \micron\ to be an AGB star,
except for those optically classified as blue objects, since
unresolved \HII\ regions lie in the same part of the IR CMD as
mass-losing AGBs. In comparison with the tracks of increasing wind
optical depth \citep[see][]{gro06,jac07}, this plot shows that the
most heavily mass-losing stars are among the most luminous AGBs. The
maximum MLRs observed in IC~1613 are in very good agreement with the
maximum MLRs seen in WLM \citep{jac07} and the LMC \citep{van99}, and
somewhat higher than the classic single-scattering MLR limit predicted
by \citet{jur84}. We also see a number of RSGs above the AGB limit,
none of which show infrared excess.

The total MLRs we derive for IC~1613 are 2.4$\times$10$^{-4}$
\solaryr\ for T$_{eff}$~=~2650 K and 85\% AMC + 15\% SiC,
4.3$\times$10$^{-4}$ for T$_{eff}$~=~3600 K and 85\% AMC + 15\% SiC,
1.0$\times$10$^{-3}$ \solaryr\ for T$_{eff}$~=~2500 K and 60\%
Silicate + 40\% Aluminum Oxide, and 8.8$\times$10$^{-4}$ for
T$_{eff}$~=~2500 and 100\% Silicate. These values are approximately a
factor of two lower than those found for WLM. Considering that a
significant fraction of the mass-loss is coming from a small number of
objects with short mass-loss timescales, the integrated MLR of a
relatively low-mass dwarf like IC~1613 could be highly time-variable
and the consistency between the MLRs of WLM and IC~1613 is notable.

\subsection{Luminosity Contribution From AGB Stars}\label{luminosity}
For point sources detected at 3.6 and 4.5 \micron\ we find a total 3.6
\micron\ flux of 72.4 mJy. 85\% of this total flux is from AGB stars
and RSGs, which is very similar to the case of WLM, where we found
79\% of the flux in point sources was from sources brighter than the
TRGB. The total flux in point sources we measure for IC~1613
constitutes 79\% of the integrated 4.5 \micron\ flux detected by
\citet{lee06}.

Following \citet{van05} and adopting an age of the current super-TRGB
stellar population of 2 Gyr we find a total stellar mass of
1.7$\times$10$^{7}$ \solar . This value is a factor of 2.6 higher than
the 6.6$\times$10$^{6}$ \solar\ estimated by \citet{lee06}. Because
the \citet{van05} method of measuring stellar mass assumes a
single-aged stellar population and is dependent on the age of that
population, our stellar mass may be over-estimated since a number of
stars above the TRGB in IC~1613 are likely RSGs younger than 2 Gyr
old.


\section{Summary of Results and Conclusions}\label{conclusions}

We present \spitzer\ {\it Space Telescope} thermal near-IR photometry
of IC~1613.  These data are compared with the optical {\it V} and {\it
I} photometry of \citet{uda01} and we find that 43\% of the IR
detected AGB stars are not detected in the optical and an additional
11\% of the AGBs are misidentified as sub-TRGB red giants. We show
that the optical incompleteness fraction is very well correlated with
the [3.6]$-$[4.5] color, which indicates that the optically undetected
AGB stars have been reddened by circumstellar material beyond the
optical completeness limits.

Our IR photometry is also compared with the narrowband optical carbon
star study of \citet{alb00} and we find their study detects 50\% of
the total AGB population, and their calculated C/M ratio is based on
18\% of this population. Further, the number of IR detected AGB stars
we find for WLM in \citet{jac07} is in excellent agreement with the
recent near-IR study of \citet{val07}, who find a C/M ratio of 0.58.

We find the total MLRs from AGB stars to be 0.2-1.0$\times$10$^{-3}$
\solaryr , depending on the assumed effective temperatures and
chemical composition of the mass-loss winds of the AGB stars. The
distribution of bolometric luminosities and MLRs of individual AGB
stars is in excellent agreement with those observed in the Magellanic
Clouds \citep{van99,van06}.

We show that the dominant contribution (85\%) to the thermal near-IR
luminosity of IC~1613 is from the AGBs and RSGs.

Finally, the properties of the AGB populations in IC~1613 (i.e., the
broadband and narrowband optical completeness fractions and MLRs) are
nearly identical to those found in \citet{jac07} for WLM. This is no
doubt due to their similar galaxy characteristics and evolutionary
histories, and is likely representative of star-forming dIs.


\acknowledgements We sincerely thank Serge Demers for providing his
narrowband photometry of WLM and IC~1613 and for extremely valuable
comments regarding narrowband optical completeness statistics. We also
thank the anonymous referee for their careful reading of the
manuscript and their valuable comments, particularly regarding the
possible detection of extrinsic carbon stars. This work is based in
part on observations made with the Spitzer Space Telescope, which is
operated by the Jet Propulsion Laboratory, California Institute of
Technology under NASA contract 1407. Support for this work was
provided by NASA through Contract Numbers 1256406 and 1215746 issued
by JPL/Caltech to the University of Minnesota. This research has made
use of NASA's Astrophysics Data System Bibliographic Services and the
NASA/IPAC Extragalactic Database (NED) which is operated by the Jet
Propulsion Laboratory, California Institute of Technology, under
contract with the National Aeronautics and Space Administration.


\clearpage


\clearpage
\begin{deluxetable}{lcc}
\tablecaption{Basic Properties of IC~1613}
\tablewidth{0pt}
\tablehead{
\colhead{Quantity} & \colhead{Value} & \colhead{Reference}}
\startdata
Right Ascension, $\alpha$(2000) & 01 04 46.4 & 1 \\
Declination, $\delta$(2000) & $+$02 08 45.9 & 1 \\
Heliocentric velocity, V$_\sun$ (km s$^{-1}$) & $-$232 km s$^{-1}$ & 1 \\
Distance, D (Mpc) & 0.73$\pm$ 0.02 & 2 \\
Morphological Type & Ir V & 3 \\
12~+~log(O/H) & 7.62 & 4 \\
Total \ion{H}{1} mass (M$_\sun$) & 6.5$\times10^7$ & 5 \\
Inclination angle (degrees) & 38 & 1 \\
Position angle (degrees) & 58 & 1 \\
Rotational velocity (km s$^{-1}$) & 25 & 1 \\
Conversion factor (pc/arcmin) & 212 & 2 \\
\enddata
\label{basic}
\tablerefs{(1) \citet{lak89}.  (2) \citet{dol01}. (3) 
\citet{vdb00}. (4) \citet{lee03}. (5) \citet{vol61}.}
\end{deluxetable}   
\clearpage



\clearpage
\begin{deluxetable}{lc}
\tablecaption{\label{stats} Detection Statistics}
\tablewidth{0pt}
\tablehead{
\colhead{\hspace{2cm}Total Point Source Detections in All Wavelengths }}
\startdata
Filter & Number \\
\hline
Both V and I & 7266     \\
3.6 \micron\ & 5211 \\
4.5 \micron\ & 3162 \\
5.8 \micron\ & 886  \\
8.0 \micron\ & 618  \\
& \\
\multicolumn{2}{c}{Detections in All Four IRAC bands, But Not V and I} \\
\hline
Object type & Number \\
\hline
Total & 183 \\
RSG (Above the AGB limit) & 3 \\ 
AGB/RSG (Above the TRGB, below the AGB limit) & 178 \\
Below the TRGB & 2 \\
& \\
\multicolumn{2}{c}{3.6 \micron\ Point Source Flux (Total Flux = 72.4 mJy)} \\
\hline
Object type & Fraction \\
\hline
Brighter than the TRGB & 85\% \\
Fainter than the TRGB & 15\% \\
& \\
\multicolumn{2}{c}{Optical Detection Fractions of IR Identified AGBs} \\
\hline
Filter & Fraction \\
\hline
V and I & 57\% \\
Detected but misidentified in V and I & 11\% \\
Narrow-band optical & 18\% \\
\enddata
\end{deluxetable}   
\clearpage


\end{document}